\documentclass[12pt,a4paper]{article}
\usepackage{cite}
\usepackage{amsmath}
\usepackage{epsfig}

\begin{document}
\textwidth=135mm
 \textheight=200mm
\begin{center}
{\bfseries How to measure the charm density in the proton
\footnote{{\small Talk given at the 20th International Symposium on Spin Physics "Spin 2012", JINR, Dubna, Russia, September 17-22, 2012.}}}
\vskip 5mm
N.~Ya.~Ivanov 
\vskip 5mm
{\small {\it Yerevan Physics Institute, Alikhanian Bros. Str. 2, 0036 Yerevan, Armenia}}
\end{center}
\vskip 5mm
\centerline{\bf Abstract}
We study two experimental ways to measure the heavy-quark content of the proton: 
using the Callan-Gross ratio $R(x,Q^2)=F_L/F_T$ and/or the azimuthal $\cos(2\varphi)$ asymmetry 
in DIS. Our approach is based on the following observations. First, the ratio $R(x,Q^2)=F_L/F_T$ and azimuthal $\cos(2\varphi)$ asymmetry in heavy-quark leptoproduction are stable, both 
parametrically and perturbatively, within pQCD. Second, both these quantities are sensitive to resummation of the mass logarithms of the type $\alpha_{s}\ln\left( Q^{2}/m^{2}\right)$. We conclude that the heavy-quark densities in the nucleon can, in principle, be determined from high-$Q^2$ data on the Callan-Gross ratio and/or the azimuthal asymmetry. In particular, the charm content of the proton can be measured in future studies at the proposed Large Hadron-Electron (LHeC) and Electron-Ion (EIC) Colliders.
\vskip 10mm
\noindent {\bf 1.~Introduction.}\ \ \  
The notion of the intrinsic charm content of the proton has been introduced about 30 years ago in Ref.~\cite{BHPS}. It was shown that, in the light-cone Fock space picture, it is natural to expect a five-quark state contribution, $\left\vert uudc\bar{c}\right\rangle$, to the proton wave function. This component has nonperturbative nature and can be generated by $gg\rightarrow c\bar{c}$ fluctuations inside the proton where the gluons are coupled to different valence quarks.

In the middle of nineties, another point of view on the charm content of the proton has been proposed in the framework of the variable-flavor-number scheme (VFNS) \cite{ACOT}.  The VFNS is an approach alternative to the traditional fixed-flavor-number scheme (FFNS) where only light degrees of freedom ($u,d,s$ and $g$) are considered as active. Within the VFNS, the
mass logarithms of the type $\alpha_{s}\ln\left( Q^{2}/m^{2}\right)$ are resummed through the all orders into a heavy quark density which evolves with $Q^{2}$ according to the standard DGLAP  evolution equation. Consequently, the charm density arises within the VFNS perturbatively via the $g\rightarrow c\bar{c}$ evolution. So, the VFNS was introduced to resum the mass logarithms and to improve thus the convergence of original pQCD series.

Presently, both nonperturbative and perturbative charm densities are widely used for a
phenomenological description of available data. In particular, practically all the recent versions of the CTEQ \cite{CTEQ} sets of PDFs are based on the VFN schemes and contain a charm density. At the same time, the key question remains open: How to measure the charm content of the proton?  The basic theoretical problem is that radiative corrections to the heavy-flavor production cross sections are large: they increase the leading order (LO) results by approximately a factor of two. Moreover, perturbative instability leads to a high 
sensitivity of the theoretical calculations to standard uncertainties in the input QCD parameters: $\mu _{F}$, $\mu _{R}$, $\Lambda_{\mathrm{QCD}}$ and PDFs. For this reason, one can 
only estimate the order of magnitude of the pQCD predictions for charm production cross sections in the entire energy range from the fixed-target experiments \cite{Mangano-N-R} to the RHIC collider \cite{R-Vogt}.

Since production cross sections are not perturbatively stable within QCD, they cannot be a good probe of the charm density. For this reason, it is of special interest to study those observables that are well-defined in pQCD. Nontrivial examples of such observables were 
proposed in Refs.~\cite{we1,we2,we3,we4}, where the azimuthal $\cos(2\varphi)$ asymmetry and Callan-Gross ratio $R(x,Q^2)=F_L/F_T$ in heavy-quark leptoproduction were analyzed.\footnote{Note also the paper \cite{Almeida-S-V}, where the perturbative stability of the QCD predictions for the charge asymmetry in top-quark hadroproduction has been observed.} It was shown that, contrary to the production cross sections, the azimuthal asymmetry \cite{we1,we2} and Callan-Gross ratio \cite{we4} in heavy flavor leptoproduction are stable within the FFNS, both parametrically and perturbatively.

In this talk, we discuss resummation of the mass logarithms of the type $\alpha_{s}\ln\left( Q^{2}/m^{2}\right)$ in leptoproduction of heavy flavors:
\begin{equation}
l(\ell )+N(p)\rightarrow l(\ell -q)+Q(p_{Q})+X[\overline{Q}](p_{X}). \label{1}
\end{equation}
The cross section of the reaction (\ref{1}) may be written as
\begin{eqnarray}
\frac{\mathrm{d}^{3}\sigma_{lN}}{\mathrm{d}x\mathrm{d}Q^{2}\mathrm{d}\varphi }&=&\frac{2\alpha^{2}_{em}}{Q^4}
\frac{y^2}{1-\varepsilon}\Bigl[ F_{T}( x,Q^{2})+\varepsilon F_{L}(x,Q^{2}) \Bigr. \nonumber \\
&&+\Bigl. \varepsilon F_{A}( x,Q^{2})\cos 2\varphi+
2\sqrt{\varepsilon(1+\varepsilon)} F_{I}( x,Q^{2})\cos \varphi\Bigr], \label{2}
\end{eqnarray}
where $F_{2}(x,Q^2)=2x(F_{T}+F_{L})$ while the quantity $\varepsilon$ measures the degree of the longitudinal polarization of the virtual photon in the Breit frame: $\varepsilon=\frac{2(1-y)}{1+(1-y)^2}$. 
The quantities $x$, $y$, and $Q^2$ are the usual Bjorken kinematic variables. 
For the azimuth $\varphi$, we use the definitions presented in Refs.~\cite{we3}.
\begin{figure}[h]
\begin{center}
\begin{tabular}{ll}
\mbox{\epsfig{file=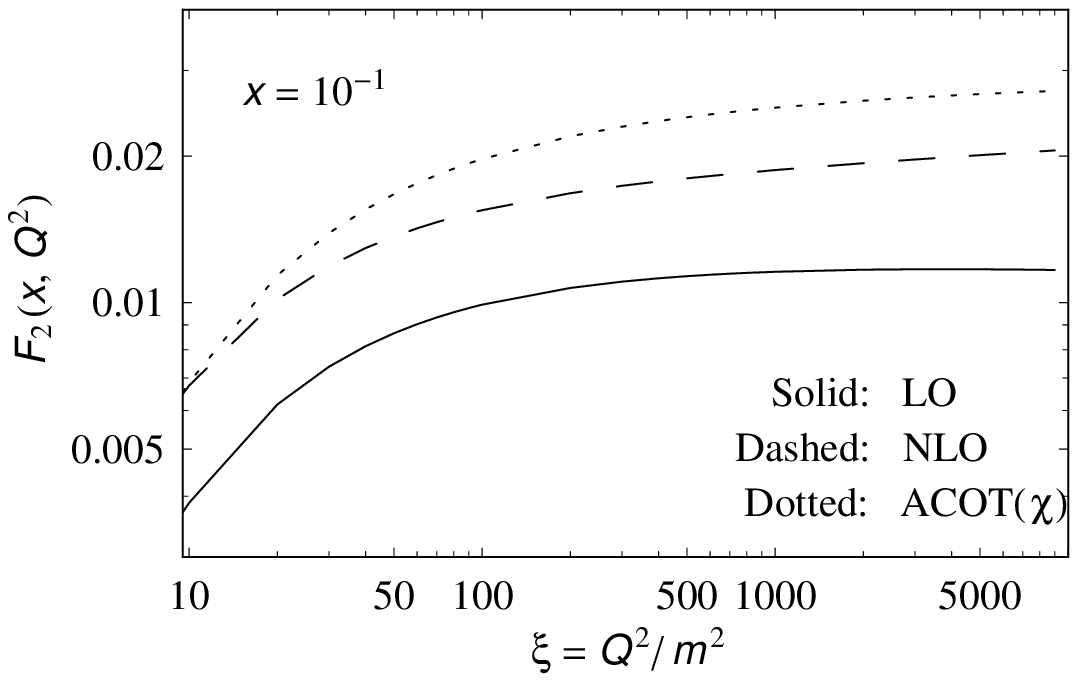,width=190pt}}
& \mbox{\epsfig{file=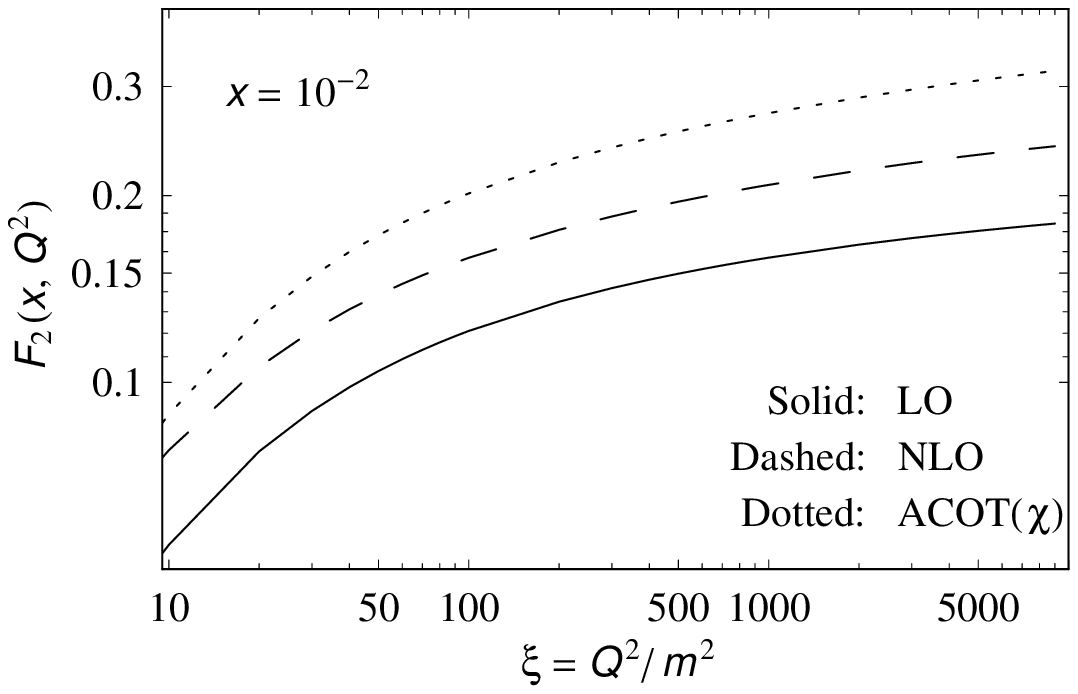,width=190pt}}
\end{tabular}
\caption{\label{Fig1}\small LO (solid lines), NLO (dashed lines) and ACOT($\chi$) (dotted curves) predictions for $F_2(x,Q^2)$ in charm leptoproduction at $x=10^{-1}$ and $10^{-2}$.}
\end{center}
\end{figure}
\begin{figure}[h]
\begin{center}
\begin{tabular}{ll}
\mbox{\epsfig{file=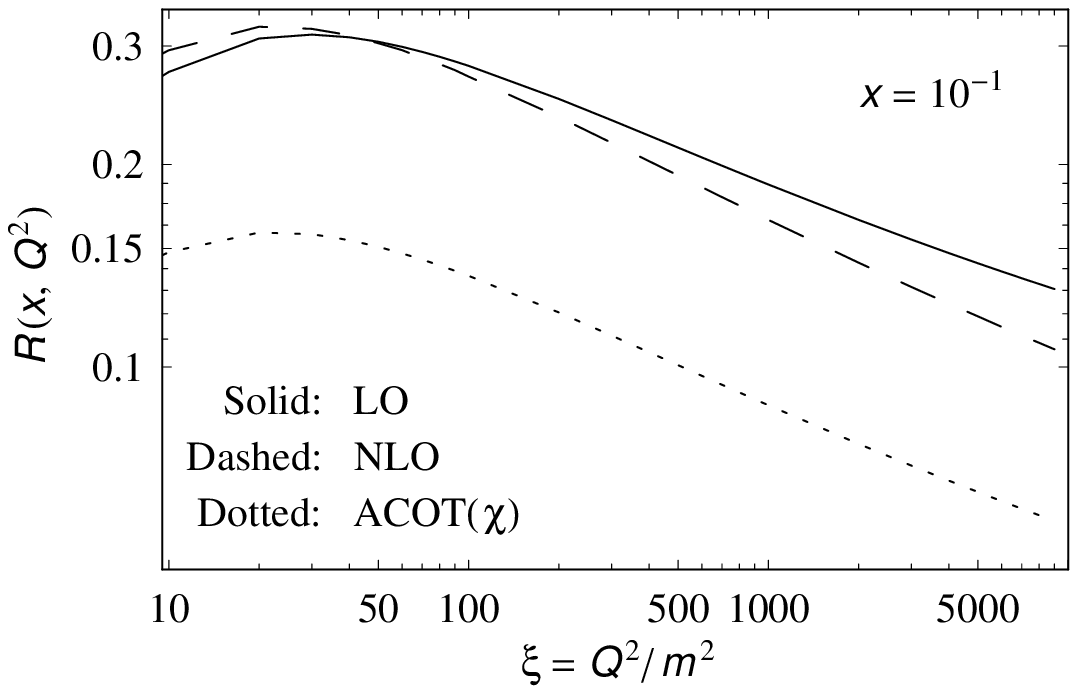,width=190pt}}
& \mbox{\epsfig{file=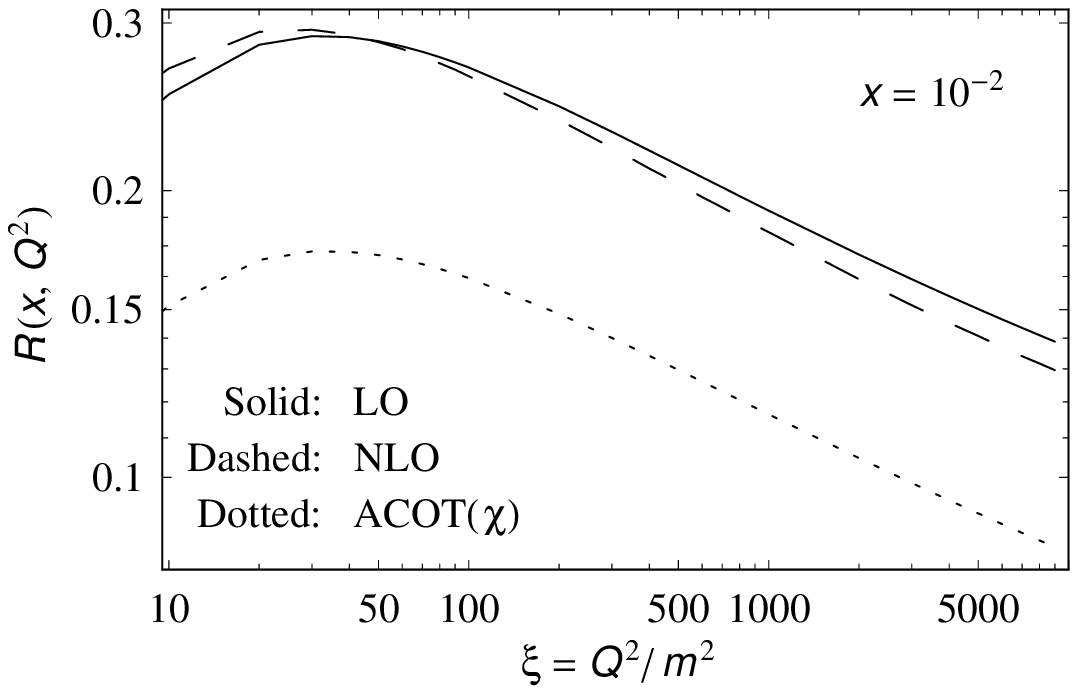,width=190pt}}
\end{tabular}
\caption{\label{Fig2}\small LO (solid lines), NLO (dashed lines) and ACOT($\chi$) (dotted curves) predictions for $R(x,Q^2)$ in charm leptoproduction at $x=10^{-1}$ and $10^{-2}$.}
\end{center}
\end{figure}

In particular, we consider resummation of the mass logarithms for the quantities $R(x,Q^2)$ and $A(x,Q^2)$ defined as
\begin{equation}\label{3}
R(x,Q^{2})=\frac{F_{L}}{F_{T}}(x,Q^{2}), \qquad \qquad A(x,Q^{2})=2x\frac{F_{A}}{F_{2}}(x,Q^{2}).
\end{equation}
\vskip 3mm
\noindent {\bf 2.~Resummation of mass logarithms.}\ \ \  
In Figs.~\ref{Fig1}  and \ref{Fig2}, we present the LO and next-to-leading order (NLO) FFNS predictions for the structure function $F_2(x,Q^2)$ and Callan-Gross ratio  $R(x,Q^2)=F_L/F_T$ in charm leptoproduction, and compare them with the corresponding ACOT($\chi$) VFNS results \cite{chi}. In our calculations, the CTEQ6M parameterization for PDFs and $m_c=1.3$~GeV for c-quark mass are used \cite{CTEQ}. The default common value for the factorization and renormalization scales is $\mu=\sqrt{4m_{c}^{2}+Q^{2}}$.

One can see from Fig.~\ref{Fig1} that both the radiative corrections and charm-initiated contributions to $F_{2}(x,Q^{2})$ are large: they increase the LO FFNS results by 
approximately a factor of two  at $x\sim 10^{-1}$ for all $Q^2$. At the same time, the relative difference between the dashed and dotted lines is not large: it does not exceed $25\%$ for $\xi=Q^2/m^2<10^{3}$. We conclude that it will be very difficult to determine the charm content of the proton using only data on $F_{2}(x,Q^{2})$ due to large radiative corrections (with corresponding theoretical uncertainties) to this quantity. 

Considering the corresponding predictions for the ratio $R(x,Q^2)$ presented in Fig.~\ref{Fig2}, 
we see that, in this case, the NLO and charm-initiated ACOT($\chi$) contributions are strongly different. In particular, the NLO corrections to $R(x,Q^2)$ are small, less than $15\%$, for $x\sim 10^{-2}$--$10^{-1}$ and $\xi<10^{4}$. This implies that large radiative contributions to the structure functions $F_{T}$ and $F_{L}$ cancel each other in the ratio $F_{L}/F_{T}$ with  good accuracy.

At the same time, the charm-initiated contributions to $R(x,Q^2)$ are large: they decrease the LO FFNS predictions by about $50\%$ practically for all values of $\xi>10$. 
This is due to the fact that resummation of the mass logarithms has different effects on
the structure functions $F_{T}(x,Q^{2})$ and $F_{L}(x,Q^{2})$. In particular, contrary to 
the transverse structure function, the longitudinal one does not contain leading mass logarithms of the type $\alpha_s\ln (Q^{2}/m^{2})$ at both LO and NLO \cite{BMSMN}. For this reason, resummation of these logarithms within the VFNS leads to increasing of the quantity $F_{T}$ but does not affect the function $F_{L}$. We conclude that the Callan-Gross ratio $R(x,Q^2)=F_L/F_T$ could be good probe of the charm density in the proton at $x\sim 10^{-2}$--$10^{-1}$.
\begin{figure}
\begin{center}
\begin{tabular}{ll}
\mbox{\epsfig{file=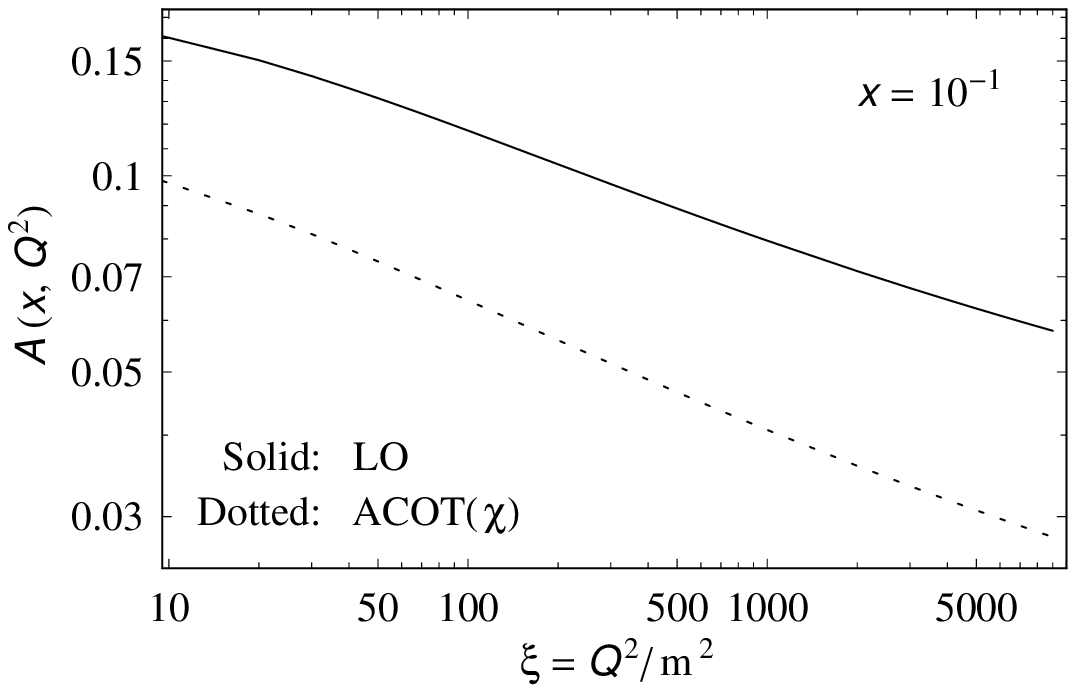,width=190pt}}
& \mbox{\epsfig{file=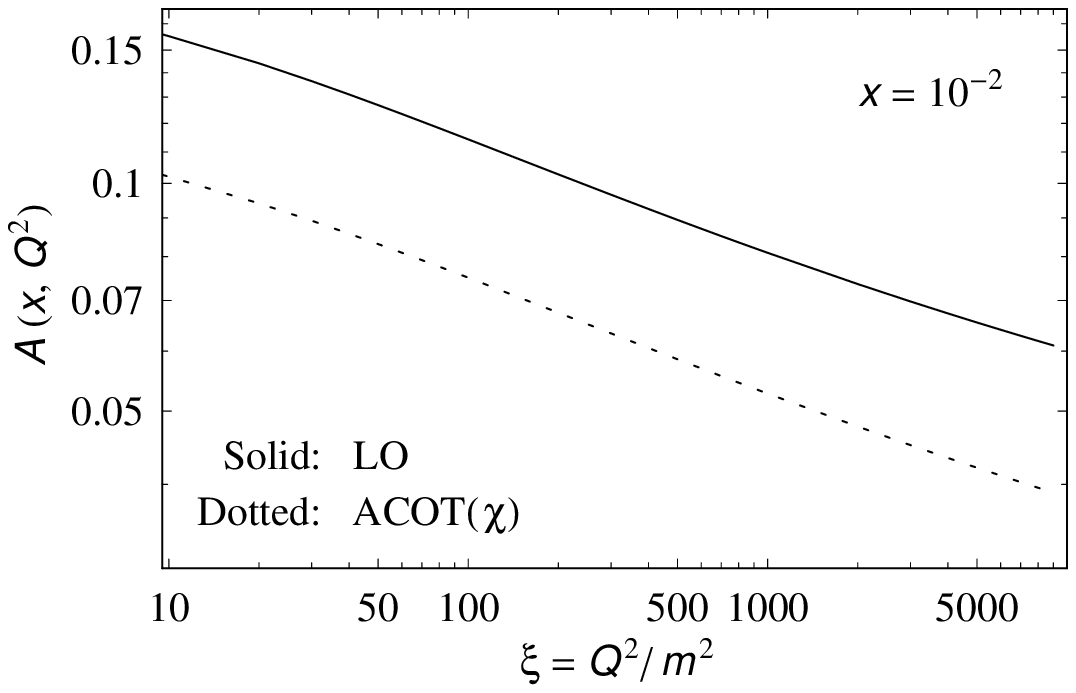,width=190pt}}
\end{tabular}
\caption{\label{Fig3}\small LO (solid lines) and ACOT($\chi$) (dotted curves) predictions for $A(x,Q^2)$ in charm leptoproduction at $x=10^{-1}$ and $10^{-2}$.}
\end{center}
\end{figure}

Fig.~\ref{Fig3} shows the LO FFNS and ACOT($\chi$) predictions for the azimuthal asymmetry $A(x,Q^2)=2xF_{A}/F_{2}$ at $x=10^{-1}$ and $10^{-2}$.
One can see from Fig.~\ref{Fig3} that the mass logarithms resummation leads to a sizeable decreasing of the LO FFNS predictions for the $\cos2\varphi$ asymmetry. In the ACOT($\chi$) scheme, the charm-initiated contribution reduces the FFNS results for $A(x,Q^{2})$ by about 
$(30$--$40)\%$ at $x\sim 10^{-2}$--$10^{-1}$. The origin of this reduction is the same as in the case of $R(x,Q^{2})$: in contrast to $F_{2}(x,Q^{2})$, the azimuth-dependent structure function $F_{A}(x,Q^{2})$ is safe in the limit $m^2\to 0$ at least at LO. 

Presently, the exact NLO predictions for the azimuth-dependent structure function $F_{A}$ are not available. However, in Ref.~\cite{we2} the radiative corrections to the $\cos2\varphi$ asymmetry have been estimated within the so-called soft-gluon approximation at $Q^2 \sim m^2$. It was demonstrated that soft-gluon corrections to both $F_{A}$ and $F_{2}$ are large but cancel each other in their ratio $A=2xF_{A}/F_{2}$ with a good accuracy. 
For this reason, it is natural to expect that the azimuthal $\cos2\varphi$ asymmetry is also a perturbatively stable quantity in wide range of variables $x$ and $Q^{2}$ within the FFNS.

We see that the impact of the mass logarithms resummation on the $\cos2\varphi$ asymmetry is essential at $x\sim 10^{-2}$--$10^{-1}$ and therefore can be tested experimentally.
\vskip 3mm
\noindent {\bf 3.~Conclusion.}\ \ \  
In the present talk, we compare the structure function $F_{2}$, Callan-Gross ratio $R=F_L/F_T$ and azimuthal asymmetry $A=2xF_{A}/F_{2}$ in charm leptoproduction as probes of the charm content of the proton. Our analysis of the radiative and charm-initiated corrections 
indicates that, in a wide kinematic range, both contributions to the structure function 
$F_{2}(x,Q^{2})$ have similar $x$ and $Q^2$ behaviors. For this reason, it will be difficult 
to estimate the charm content of the proton using only data on $F_{2}(x,Q^{2})$.

The situation with the Callan-Gross ratio and azimuthal asymmetry looks more promising. 
Our analysis shows that resummation of the mass logarithms leads to reduction of the FFNS predictions for $R(x,Q^2)$ and $A(x,Q^2)$ by $(30$--$50)\%$ at $x\sim 10^{-2}$--$10^{-1}$ and $Q^2\gg m^2$. Taking into account the perturbative stability of the Callan-Gross ratio and 
azimuthal asymmetry within the FFNS, we conclude that it will, in principle, be possible to determine the charm density in the proton from future high-$Q^2$ data on $R=F_L/F_T$ and $A=2xF_{A}/F_{2}$. 
\vskip 3mm
\noindent {\emph{Acknowledgments.}}\ \ \
The author is grateful to Organizing Committee of the "SPIN 2012" Symposium for invitation and support. We thank S.~J.~Brodsky, A.~V.~Efremov, A.~V.~Kotikov, A.~B.~Kniehl, E.~Leader and C.~Weiss for useful discussions. This work is supported in part by the State Committee of Science of RA, grant 11-1C015.

\end{document}